\documentclass[aps,prd,twocolumn,superscriptaddress,groupedaddress]{revtex4-1}  % for review and submission
\usepackage{graphicx}
\usepackage[usenames]{color}
\usepackage{amsmath}
\usepackage{amssymb}
\usepackage{amsthm}
\usepackage{bbold}
\usepackage{float}

\newcommand{\bra}[1]{\ensuremath{\langle #1 |}}
\newcommand{\ket}[1]{\ensuremath{| #1 \rangle}}

\newcommand{\kd}{K{\"a}hler-Dirac }
\renewcommand{\dag}{\dagger}
\DeclareMathOperator{\bound}{d}

\newcommand*{\CPT}{\raise0.4ex\hbox{$\chi$}PT}
\newcommand*{\chpt}{\raise0.4ex\hbox{$\chi$}PT}
\newcommand*{\schpt}{S\raise0.4ex\hbox{$\chi$}PT}

\def\eqref#1{{(\ref{#1})}}

\def\bar{\overline}
\def\hat{\widehat}

\def\bea{\begin{eqnarray}}
\def\eea{\end{eqnarray}}

\begin{document}
\title{\kd fermions on Euclidean dynamical triangulations}
\author{Simon Catterall}
\author{Jack Laiho}
\author{Judah Unmuth-Yockey}
\affiliation{Department of Physics, Syracuse University, Syracuse, NY 13244 USA}
\date{\today}

\begin{abstract}
We study \kd fermions on Euclidean dynamical triangulations.  This fermion formulation furnishes a
natural extension of staggered fermions to random geometries without requring
vielbeins and spin connections. We work in the quenched approximation where
the geometry is allowed to fluctuate but there is no back-reaction from the matter on the geometry. By examining
the eigenvalue spectrum and the masses of scalar mesons we find evidence for a four fold
degeneracy in the fermion spectrum in the large volume, continuum limit. It is natural to associate this degeneracy
with the well known equivalence in continuum flat space between the \kd fermion and four copies of a Dirac fermion. Lattice effects then lift this
degeneracy in a manner similar to staggered fermions on regular lattices.
The evidence that these discretization effects vanish in the continuum limit suggests both that lattice
continuum \kd fermions are recovered at that point, and that this limit truly corresponds to smooth continuum geometries.
One additional advantage of the \kd action is that it respects
an exact $U(1)$ symmetry on any random triangulation. This $U(1)$ symmetry is related to continuum chiral
symmetry. By examining fermion bilinear condensates we find strong evidence
that this $U(1)$ symmetry is {\it not} spontaneously broken in the model at order the Planck scale.  This is a necessary requirement if 
models based on dynamical triangulations are to provide a valid ultraviolet complete formulation of quantum gravity.
\end{abstract}

\maketitle

\section{Introduction}

Quantum gravity remains one of the outstanding challenges in theoretical physics.  One approach to obtaining a consistent, predictive theory is the asymptotic safety scenario of Weinberg, where the theory is effectively renormalizable when formulated nonperturbatively \cite{Weinberg:1980gg}.  It is already well-known that gravity is not perturbatively renormalizable \cite{'tHooft:1974bx, Goroff:1985th}, so the investigation of asymptotic safety rests on the hope that the theory is strongly coupled at the ultraviolet fixed point, a situation that is difficult to investigate using standard perturbative field theory methods.  Despite this, there are a number of results that suggest that gravity is asymptotically safe, coming mainly from either lattice \cite{Ambjorn:2005qt,Ambjorn:1998xu,Ambjorn:2008wc}, or from the functional renormalization group \cite{Reuter:2001ag,Lauscher:2001ya,Litim:2003vp,Codello:2007bd,Codello:2008vh,Benedetti:2009rx}.  This work follows up on Ref.~\cite{Laiho:2016nlp} in focusing on Euclidean dynamical triangulations (EDT), which is one of the original attempts to formulate quantum gravity using lattice methods \cite{Ambjorn:1991pq, Agishtein:1991cv,Catterall:1994pg}.

EDT is a path integral formulation of gravity that involves a sum over all geometries, weighted by the Einstein-Hilbert action.  The geometries are approximated by piecewise continuous assemblies of 4-dimensional building blocks called 4-simplices, which are the four dimensional analogs of triangles.  If a continuous phase transition can be found somewhere in the phase diagram, the diverging correlation length associated with a continuous phase transition would allow one to take a continuum limit.  If in addition to this, classical general relativity in four dimensions emerges in the appropriate long distance limit, then dynamical triangulations would be a candidate formulation for an ultraviolet complete theory of quantum gravity. 

In Ref.~\cite{Laiho:2016nlp}, which involved a subset of the authors of this work, we presented evidence that EDT supplemented by an ultralocal measure term \cite{Bruegmann:1992jk} can recover semiclassical geometries that are four dimensional within measurement errors, and that this model has a continuum limit.  The evidence for semiclassical physics is the following:  The global Hausdorff dimension is close to four, as seen by finite volume scaling; the expectation value of the geometries resembles Euclidean de Sitter space in four dimensions, and the agreement with the classical solution gets better as the continuum limit is approached; the spectral dimension, which is a fractal dimension defined by a diffusion process, varies with distance scale and approaches a value close to four at long distance scales.  This variation of the spectral dimension with distance scale is seen in other approaches to quantum gravity \cite{Ambjorn:2005db, Lauscher:2005qz, Carlip:2017eud}.  The recovery of the physical theory appears to require a fine-tuning of the bare lattice parameters.  In fact, the agreement with the classical theory is not so good at long distances, but gets better as one makes the lattice spacing smaller.  Both of these are features of a theory where a symmetry is broken by the lattice regulator.  This implies that discretization errors are actually smallest in a window between the largest and smallest distance scales on a given lattice ensemble, though this situation improves as we take the continuum limit.  We must keep this in mind when choosing fit ranges for our matter correlation functions.  It was argued in Ref.~\cite{Laiho:2016nlp} that the symmetry in the case of our calculations is continuum diffeomorphism invariance.  Regardless of this interpretation, the numerical evidence for the emergence of semiclassical physics argues for further investigation of the model.

One important way to test this approach is to study the effects of dynamical triangulations on the matter sector, and in this work we focus on fermionic matter.  Given the fractal nature of the geometries at short distances, it is not a forgone conclusion that we can recover continuum-like fermions in the long distance limit.  Furthermore, it is important that strongly coupled gravity does not behave like quantum chromodynamics (QCD) in that quantum fluctuations in the gravity sector should not induce chiral symmetry breaking in the fermion sector.  In QCD, spontaneous chiral symmetry breaking leads to fermion bound states with masses of order the QCD scale, except for the pions, which are the pseudo-Goldstone bosons of the spontaneous chiral symmetry breaking.  In the Standard Model, chiral symmetry protects fermion masses from additive renormalization, so that light fermions are technically natural.  We do not want quantum gravity to lead to spontaneous chiral symmetry breaking induced at the Planck scale, since this would cause fermion bound states to also acquire masses of order the Planck scale. Notice that this is not just an academic point: early work
with Dirac fermions on quenched random lattices saw precisely this phenomenon and one might worry that a similar
phenomenon would occur on DT geometries \cite{Espriu1986,Catterall:1988an}. However, in this paper we will show strong evidence that this is not the case for
\kd fermions. At least part of the underlying reason for 
this stems from the existence of an exact 
$U(1)$ symmetry for \kd fermions which generalizes the usual $U(1)$ symmetry of staggered quarks in QCD and ensures that fermion masses
undergo multiplicative renormalization and depend only weakly on the cut-off scale.
 
As a further step in testing the EDT model we investigate adding fermions to the simulations.  Calculations involving scalars \cite{Ambjorn:1993ta} and gauge fields \cite{Ambjorn:1994ev, Bilke:1998vj} have been done on dynamically triangulated lattices some time ago, but fermions have only been considered in two dimensions \cite{Burda:1999av}.  In this work we study fermions in four dimensions using the K\"ahler-Dirac formulation \cite{Kahler:1962}.  This is a natural approach to fermions on dynamically triangulated geometries because of the precise correspondence between the continuum and lattice formulations and because the construction does not require the addition of new degrees of
freedom like the vielbein and spin connection. In flat, four dimensional space-time continuum K\"ahler-Dirac fermions are equivalent to four copies of Dirac fermions, but they differ in the presence of curvature.  Thus, if our geometries are sufficiently smooth, then in the large volume, vanishing curvature limit, we might expect to recover four copies of Dirac fermions.  We will see that there is evidence supporting this, but only in the continuum limit.  At finite lattice spacing we see discretization effects that are reminiscent of staggered fermions in lattice gauge theories \cite{Bazavov:2009bb}.  This may not be surprising, given the close relationship between staggered fermions and K\"ahler-Dirac fermions in gauge theory on a regular hypercubic lattice \cite{Banks1982, Susskind:1976jm}.  We find evidence that these discretization effects vanish in the continuum limit, bolstering the case that we can include fermions with the expected continuum properties on dynamical triangulations, and perhaps more importantly, that a continuum limit exists at all.  We also find evidence that the analog of chiral symmetry is not spontaneously broken in our model, as indicated by the vanishing of the chiral condensate in the chiral limit when the infinite volume limit is also taken.  This is a necessary requirement for a phenomenologically viable theory of quantum gravity.

This paper is organized as follows:  Section~\ref{sec:latsim} reviews the EDT formulation and the lattice simulations that were carried out in Ref.~\cite{Laiho:2016nlp}.  Section~\ref{sec:kdferm} reviews the K\"ahler-Dirac formulation, both in the continuum and on the lattice.  Section~\ref{sec:ddel} discusses the construction of the exterior derivatives on the lattice.  Section~\ref{sec:results} presents the results of the numerical simulations, including a study of the eigenvalue spectrum and fermion bound states, as well as fermion condensates.  Finally, we conclude in Section~\ref{sec:conc}.

\section{Lattice simulations}
\label{sec:latsim}

\subsection{The model}

In Euclidean quantum gravity the partition function is formally given by a path integral sum over all geometries weighted by the Einstein-Hilbert action and the action associated with the matter sector
\bea\label{eq:part}  Z_E = \int {\cal D}[g]{\cal D}[\overline{\omega}]{\cal D}[\omega] e^{-S_{EH}[g] - S_{KD}[\overline{\omega}, \omega]},
\eea
where the Euclidean Einstein-Hilbert action is
\bea  \label{eq:ERcont} S_{EH} =  -\frac{1}{16 \pi G}\int d^4x \sqrt{g} (R - 2\Lambda),
\eea
with $R$ the Ricci curvature scalar, $g$ the determinant of the metric tensor, $G$ Newton's constant, $\Lambda$ the cosmological constant, and we choose the matter action $S_{KD}$ to be the K\"ahler-Dirac action
\bea  S_{KD} = \int d^4x \sqrt{g} \  \overline{\omega}(\bound - \, \delta + m_0)\omega.
\eea
The K\"ahler-Dirac fields, $\omega$ consist of a collection of Grassmann valued $p$-form fields which are coupled
through the exterior derivative $\bound$ and its adjoint $\delta$ with $m_0$ the mass. The equivalence of
this action to (four copies of) the usual Dirac action in flat space is shown in \cite{Banks1982,kdferm2018}. Notice that this
action is only sensitive to the geometry via the measure $\sqrt{g}$ and the hidden factors of the metric needed to
define the adjoint $\delta$. 
    
In dynamical triangulations, the path integral is formulated directly as a sum over geometries, without the need for gauge fixing or the introduction of a metric.  The dynamical triangulations approach is based on the conjecture that the path integral for Euclidean gravity without matter is given by the partition function \cite{Ambjorn:1991pq, Bilke:1998vj}
\bea\label{eq:Z} Z_E = \sum_T \frac{1}{C_T}\left[\prod_{j=1}^{N_2}{\cal O}(t_j)^\beta\right]e^{-S_{ER}}
\eea
where $C_T$ is a symmetry factor that divides out the number of equivalent ways of labeling the vertices in the triangulation $T$.  The term in brackets in Eq.~(\ref{eq:Z}) is a nonuniform measure term, where the product is over all two-simplices (triangles), and ${\cal O}(t_j)$ is the order of triangle $j$, i.e. the number of four-simplices to which the triangle belongs.  In Ref.~\cite{Laiho:2016nlp} a subset of us found that the parameter $\beta$ must be fine-tuned in order to obtain ensembles with semiclassical properties in four dimensions.  In this work we neglect the contribution of $S_{KD}$ in Eq.~(\ref{eq:Z}) in the Boltzmann weight when generating the geometries.  This is an uncontrolled approximation commonly referred to in lattice field theory as the quenched approximation, but it allows us to use the existing ensembles that were generated previously and described in detail in Ref.~\cite{Laiho:2016nlp}.  Even quenched, the properties of the valence fermions and their interactions can be studied, and a number of qualitative features are expected to be retained.

The discretized version of the Einstein-Hilbert action in four dimensions is the Einstein-Regge action \cite{Regge:1961px}
  \begin{equation} \label{eq:GeneralEinstein-ReggeAction}
S_{E}=-\kappa\sum_{j=1}^{N_2} V_{2}\delta_j+\lambda\sum_{j=1}^{N_4} V_{4},
\end{equation}
  \noindent where $\delta_j=2\pi-{\cal O}(t_j)\arccos(1/4)$ is the deficit angle around a triangular hinge $t_j$, with ${\cal O}(t_j)$ the number of four-simplices meeting at the hinge, $\kappa=\left(8\pi G \right)^{-1}$, $\lambda=\kappa\Lambda$, and the volume of a $d$-simplex is 
\begin{equation} \label{eq:SimplexVolume}
V_{d}=\frac{\sqrt{d+1}}{d!\sqrt{2^{d}}}a^d,
\end{equation}
\noindent where the equilateral $d$-simplex has a side of length $a$.  After performing the sums in Eq.~(\ref{eq:GeneralEinstein-ReggeAction}) one finds
\begin{equation}\label{eq:DiscAction}
S_{E}=-\frac{\sqrt{3}}{2}\pi\kappa N_{2}+N_{4}\left(\kappa\frac{5\sqrt{3}}{2}\mbox{arccos}\frac{1}{4}+\frac{\sqrt{5}}{96}\lambda\right),
\end{equation}
where $N_i$ is the number of simplices of dimension $i$.  We can rewrite the Einstein-Regge action in the simple form 
\bea\label{eq:ER}  S_{ER}=-\kappa_2 N_2+\kappa_4N_4,
\eea
where we have introduced the couplings $\kappa_2$ and $\kappa_4$, which are more convenient to work with in the simulations.

The generation of the lattice ensembles is described in detail in Ref.~\cite{Laiho:2016nlp}, and we make use of the same saved lattices that were used in that work.  We briefly review the properties of those lattices here.  
The lattice geometries are constructed by gluing together four-simplices along their ($4-1$)-dimensional faces.  The four-simplices are equilateral, with constant edge length $a$, and the dynamics is encoded in the connectivity of the simplices.  We sum over a set of degenerate triangulations, where the usual combinatorial manifold constraints are relaxed \cite{Bilke:1998bn}.  Thus, distinct four-simplices may share the same 5 distinct vertex labels, and the neighbors of a given 4-simplex are not necessarily unique; these are both conditions that would violate the combinatorial manifold constraints.  The advantage of using degenerate triangulations is that it leads to a factor of $\sim$10 reduction in finite-size effects compared to combinatorial triangulations \cite{Bilke:1998bn}.  Evidence for the existence of a continuous phase transition and a continuum limit was presented in Ref.~\cite{Laiho:2016nlp} for degenerate triangulations.  If this continuum limit really exists, it is likely that by universality the sum over combinatorial triangulations would realize the same continuum theory.  The use of degenerate triangulations introduces a complication when formulating K\"ahler-Dirac fermions on our lattice geometries, since the formulation requires that the geometries be orientable.  We show that the class of degenerate triangulations retains this property and that the expected relations for the lattice analogues of the operators $\bound$ and $\delta$ are still satisfied.

\subsection{Simulation details}

The numerical methods used to evaluate the partition function in Eq.~(\ref{eq:Z}) are by now well established \cite{Ambjorn:1997di}.  The standard (scalar) algorithm to perform the Monte Carlo integration of the dynamical triangulations partition function consists of an ergodic set of local moves, known as the Pachner moves, which are used to update the geometries \cite{Agishtein:1991cv, Gross:1991je,Catterall:1994sf}, and a Metropolis step, which is used to accept or reject the proposed move.  The lattices in this work were generated using a parallel variant of the standard algorithm, called parallel rejection.  This algorithm was tested by demonstrating that it gives identical results, configuration by configuration, to the scalar algorithm.  Parallel rejection takes advantage of, and partially compensates for, the low acceptance of the Metropolis step in the range of couplings that were simulated and is described in more detail in Ref.~\cite{Laiho:2016nlp}.  

The sum over geometries is restricted to fixed global topology $S^4$.  In order to restrict the geometries to $S^4$ it is sufficient to start from the minimal four-sphere at the beginning of the Monte Carlo evolution, since the local moves are topology preserving.  As is standard in such simulations, we tune $\kappa_4$ to its critical value in order to take the infinite lattice-volume limit.  This amounts to a tuning of the bare cosmological constant, although this is not in itself sufficient to recover a small renormalized cosmological constant, since it only allows one to take the infinite lattice-volume limit, not the infinite physical-volume limit.  
In four dimensions the update moves are only ergodic if the lattice four-volume is allowed to vary.  However, it is convenient to keep $N_4$ approximately fixed.  We work at (nearly) fixed four-volume by including a term in the action $\delta\lambda |N_4^f-N_4|$ to keep the four-volume close to a fiducial value $N_4^f$.  This does not alter the action at values of $N_4=N_4^f$, but serves to keep the volume fluctuations about $N_4^f$ from growing too large for practical simulations.  The value of $\delta \lambda$ was chosen to be 0.04, which was found to be small enough to not lead to significant effects \cite{Laiho:2016nlp}.  

\begin{figure}
\begin{center}
\includegraphics[scale=.55]{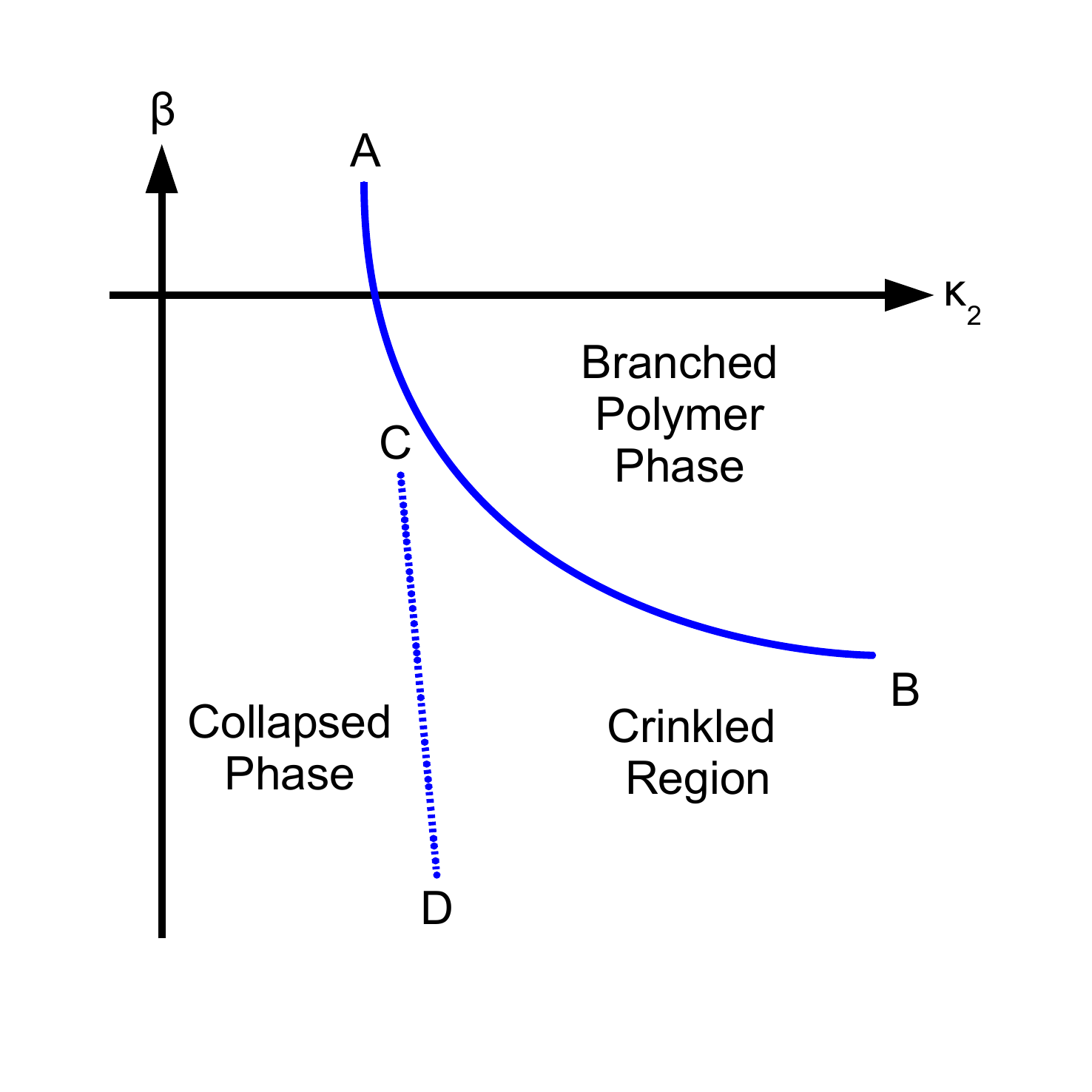}
\vspace{-3mm}
\caption{Schematic of the phase diagram as a function of $\kappa_2$ and $\beta$. \label{fig:phase1}}
\end{center}
\end{figure}

The phase diagram of this model is shown in Fig.~\ref{fig:phase1}.  Line $AB$ is a first order phase transition line separating the branched polymer phase and the collapsed phase.  Neither of these phases looks particularly semiclassical, with the branched polymer phase having a Hausdorff dimension of 2 and the collapsed phase having a large, possibly infinite, dimension.  The crinkled region is connected to the collapsed phase by what appears to be an analytic cross-over, and it shares the properties of the collapsed phase, except that it appears to have especially large finite-size effects \cite{Coumbe:2014nea}.  A similar phase diagram was seen in Ref.~\cite{Ambjorn:2013eha} using combinatorial triangulations.  It was shown by a subset of us in Ref.~\cite{Laiho:2016nlp} that close to the phase transition line the lattices have semiclassical properties, with a Hausdorff dimension close to 4 and the emergence of de~Sitter-like geometries.  In order to obtain  classical behavior one requires a fine-tuning to the transition line, though agreement with classical results also requires that one take the continuum limit.  A continuum limit appears to exist and can be approached by following the transition line to large (possibly infinite) $\kappa_2$ values.

\begin{table}
\begin{center}
\begin{tabular}{ccccc}
\hline \hline
\ \ $a_{\rm rel}$ \ \ & \ \ $\beta$ \ \ \ & \ $\kappa_2$ \ & \ \ \ \ \ $N_4$ \ \ & \ \ \ Number of configurations \\
\hline
1.47(10) & 1.5 & 0.5886 & \ \ 4000 & 367  \\
1.47(10) & 1.5 & 0.612  & \ \ 8000 & 145 \\
1 & 0.0 & 1.669 &         \ \ 4000 & 575  \\
1 & 0.0 & 1.7024 &     \ \  8000 & 489  \\
1 & 0.0 & 1.7325 &     \ \ 16000 & 501  \\
0.81(6) & $-0.6$ & 2.45 & \ \ 4000 & 414  \\
0.72(5) & $-0.8$ & 3.0 &  \ \ 8000 & 1486  \\
\hline
\end{tabular}
\caption{The parameters of the ensembles used in this work.  The first column shows the relative lattice spacing as determined in Ref~\cite{Laiho:2016nlp}, with the ensembles at $\beta=0$ serving as the fiducial lattice spacing.  The quoted error is a systematic error associated with finite-volume effects.  The second column is the value of $\beta$, the third is the value of $\kappa_2$, the fourth is the number of four-simplices in the simulation, and the fifth is the number of configurations sampled.}
\label{tab:ensembles}
\end{center}
\end{table}

Table~\ref{tab:ensembles} shows the parameters for the EDT ensembles that we have used for fermion calculations in the current work.  We consider multiple volumes at $\beta=0$ in order to study the finite-size scaling in this region, and we consider many different points along the first-order transition line in order to study the lattice spacing dependence of valance fermion observables.  The relative lattice spacings of these ensembles was determined in Ref.~\cite{Laiho:2016nlp} by looking at the return probability of a diffusion process.

\section{\kd fermions}
\label{sec:kdferm}

\subsection{In the continuum}
\kd fermions naturally arise when taking the square root of the Laplacian operator
written in the language of exterior derivatives. In concrete terms
we form the \kd operator
\begin{equation}
	\label{eq:kdop}
	K = \bound - \, \delta.
\end{equation}
where $\bound$ is the exterior derivative and $\delta$ its adjoint.
Using the fact that $\bound^{2} =\delta^2= 0$, we see that
\begin{equation}
	K K^{\dagger} = -(\bound - \, \delta)^{2}
    = (\bound \delta + \delta \bound) = -\Box
\end{equation}
where $\Box$ is the (Euclidean) Laplacian.  A natural equation for fermions is then the \kd equation:
\begin{equation}
\label{eq:kdeq}
\left(\bound-\,\delta +m_0\right)\omega=0.
\end{equation}
Notice that the \kd field $\omega$ must necessarily correspond to a collection of differential forms. In order to build the relationship between \kd fermions and Dirac fermions we will compare the action of \eqref{eq:kdop} on differential forms with that of the Dirac operator acting on $4\times 4$ matrices.

Consider two objects, one a matrix formed from combinations of gamma matrices, and the other a collection of differential forms \cite{Banks1982,Rabin1982}:
\begin{align}
	\label{eq:forms}
	\omega &= A + A_{\mu}dx^{\mu} + \frac{1}{2}A_{\mu \nu} dx^{\mu}\wedge dx^{\nu} + \ldots \\
    \label{eq:gammas}
    \Psi &= A + A_{\mu}\gamma^{\mu} + \frac{1}{2}A_{\mu \nu} \gamma^{\mu} \gamma^{\nu} + \ldots
\end{align}
The action of \eqref{eq:kdop} on \eqref{eq:forms} is to raise and lower forms: $\bound$ acts to raise forms up by one, while $\delta$ acts to lower forms by one.  By applying \eqref{eq:kdop} to \eqref{eq:forms} and $\gamma^{\mu}\partial_{\mu}$ to \eqref{eq:gammas} in parallel, one can see that these two operators act identically on their respective objects.  It remains to deduce the relationship between
\begin{equation}
	\label{eq:psidirac}
	(\gamma^{\mu}\partial_{\mu} + m_0)\Psi = 0,
\end{equation}
and the original Dirac equation.  In fact, $\Psi$ represents four copies of a Dirac spinor, one for each column of the $4\times 4$ matrix $\Psi$.  One can see this more explicitly by writing an original Dirac spinor as a single column of a $4\times 4$ matrix where the other columns are set to zero, and since the gamma matrices are a basis for this matrix, this matrix admits a form like Eq.~\eqref{eq:gammas}.  Therefore, \eqref{eq:psidirac} corresponds to four copies of the Dirac equation.

The discussion so far has been in flat space-time; however,  Eq.~\eqref{eq:kdeq} is valid in any metric 
in contrast to the usual Dirac equation. Nevertheless, it is guaranteed that in the limit of small curvature Eq.~\eqref{eq:kdeq} is equivalent to four copies of the Dirac equation.

\subsection{On the lattice}

A remarkable advantage of \kd fermions is their immediate translation to triangulated spaces \cite{Rabin1982,montvay_munster_1994}.  Since the \kd formulation contains no spinors
there is no need to introduce a spin-connection or vielbein. Instead one can simply use known results from homology theory \cite{Nash1983, Rabin1982, montvay_munster_1994} to map the exterior derivatives to operators that act naturally on discrete spaces:
\begin{align}
	\bound &\mapsto \bar{\bound} \\
    \delta &\mapsto \bar{\delta}.
\end{align}
$\bar{\bound}$ is the co-boundary operator and $\bar{\delta} = \bar{\bound}^{T}$ is the boundary operator. These operators have actions on fields defined on $p$-simplices ($p$-cochains) which are analogous to
the action of $\bound$ and $\delta$ on continuum differential forms. In a  combinatorial triangulation each $p$-simplex
is specified by an ordered list of its vertices. The boundary operator $\bar\delta$ acts on such
a $p$-simplex to product a list of its boundary $(p-1)$-simplices weighted by signs denoting their orientation
within that $p$-simplex via the relation
\begin{equation}
    \bar{\delta}_{p} [v_0, \ldots v_{p}] = \sum_{i=0}^{p} (-1)^i [v_{0}, \ldots \hat{v}_{i} \ldots v_{p}]
\end{equation}
with the $\hat{v}_{i}$ indicating the deletion of the $i$\textsuperscript{th} vertex.
When applied to a $p$-cochain this operation yields oriented sums of boundary $(p-1)$-simplex fields for each $p$-simplex and constitutes a map from $(p-1)$-cochains to $p$-cochains.
Conversely the fundamental action of $\bar{\bound}$ is to extract the co-boundary of a given $p$-simplex, that is, to furnish a oriented list of the $(p+1)$-simplices which contain that $p$-simplex.  Applied to a $p$-cochain
it yields oriented sums of co-boundary $(p+1)$-simplex fields for each $p$-simplex and yields a map from
$(p+1)$-simplices to $p$-simplices.

It is crucially important in this construction that the triangulation be oriented. This means that in $d$ dimensions
every $(d-1)$-simplex is assigned an opposite orientation in the two $d$-simplices that contain it. It is also important
to understand how this prescription is applied to degenerate triangulations which will be covered in the following
section.

Using these operators the lattice \kd equation can be written down straightforwardly as
\begin{equation}
	(\bar{\bound} - \bar{\delta} + m_0)\bar{\omega} = 0
\end{equation}
where $\bar{\omega}$ is the collection of $p$-cochains for that lattice.  Note that, just as above, the square of this lattice operator gives the lattice Laplacian
\begin{equation}
	\Delta = (\bar{\bound} - \bar{\delta})^{2} = -(\bar{\bound} \bar{\delta} + \bar{\delta} \bar{\bound}).
\end{equation}
This matrix is block diagonal with five blocks, one for each $p$-simplex.  Therefore, the square
of the \kd operator yields lattice Laplacians for not only the 0- and 4-simplices which have a natural
interpretation in terms of the direct and dual lattices but also for all intermediate rank simplices.  In the next section we will construct these lattice operators explicitly for triangulated spaces.

\section{Constructing $\bar{\bound}$ and $\bar{\delta}$}
\label{sec:ddel}

In our work we have utilized degenerate Euclidean triangulations of a sphere in four dimensions\footnote{For recent work using \kd fermions on combinatorial triangulations see \cite{kdferm2018}}. From the
previous arguments we require that all such triangulations must be oriented. For combinatorial
triangulations any $p$-simplex is uniquely specified by its vertices. This is not the case for degenerate
triangulations and additional labels are needed to uniquely specify the $p$-simplex. Nevertheless, once these
additional labels are employed the construction of $\bar\bound$ and $\bar\delta$ follow the same pattern
discussed above.

To ensure that the surface is oriented, first the vertex labels of each 4-simplex are ordered.  Then, adjacent faces of every pair of neighboring 4-simplices need to appear with opposite orientation or parity; that is, when the boundary operator operates on two adjacent 4-simplices, the common faces they share must appear with opposite signs.  
We can accomplish this orienting by systematically going through the lattice, one 4-simplex at a time, and ensuring that each face appears with the correct parity.
To do this, consider the dual lattice.
In the dual, there is a vertex for each 4-simplex, and five edges leaving each vertex, each representing
a 3-simplex (tetrahedron).  In the case of degenerate triangulations, multiple connections between two dual 
vertices (4-simplices) are allowed, however there cannot be more than four connections between any two
dual vertices since five connections would imply that such pairs of 4-simplices would be completely disconnected from the rest of the lattice.
To go through the lattice systematically we find a spanning tree in the dual representation such that we visit each 4-simplex once.  If $N_4$ is the number of 4-simplices, then the spanning tree contains  $N_{4}-1$ edges and this tree is the path we follow to orient the lattice.

\begin{figure}[t]
	\includegraphics[width=8.6cm]{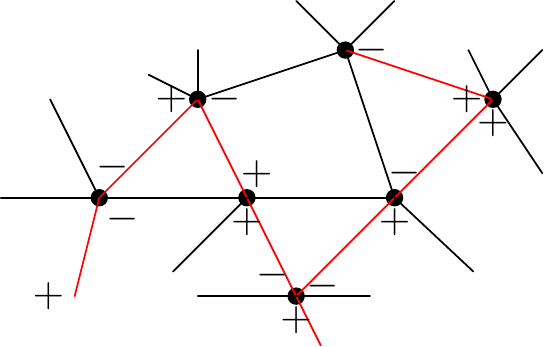}
    \caption{A section of a lattice showing oriented 4-simplices.  It is implied that at the end of each edge is another 4-simplex even if not drawn.  Here the faces of adjacent 4-simplices must have opposite parity.  Once the parity of a single face in a 4-simplex in known, all other parities for the faces in the 4-simplex are known (after fixing a vertex ordering).  A spanning tree is shown in red.}
    \label{fig:orient}
\end{figure}

To orient the lattice we first choose an ordered 4-simplex and then, using the boundary operator, assign a sign to each of the faces of the 4-simplex.  At least one of the faces of this 4-simplex is in the spanning tree.  We follow one of these edges to an adjacent 4-simplex, and using the ordered vertices of the new 4-simplex, compute the parity of its faces.  If the parity of the traversed face appears with opposite parity from whence we came, it is already correctly oriented.  Otherwise, we multiply \emph{all} of the faces in that simplex by $-1$ to give it the correct orientation.  Next we choose another edge in the tree that can be traversed and repeat the above procedure until no more 4-simplices remain to be visited.  After completing this procedure, all 4-simplices will have been visited and given an orientation based on their neighboring 4-simplices.  An example of an oriented section of the lattice is shown in Fig.~\ref{fig:orient}.
Note that the sign of a single face in a 4-simplex is enough to determine the parity of all the faces of the simplex.

As a demonstration consider the matrix which has for every 3-simplex the mapping to its oriented co-boundary i.e
an oriented list of 4-simplices.  Call this matrix $\bar{\bound}_{4}$, its shape is $N_{3} \times N_{4}$.  Then each column has five nonzero entries, one for each tetrahedron, whose matrix element is either a $1$ or $-1$ depending on the simplex orientation.  Simultaneously, each row has only two nonzero entries, and one of these is necessarily $1$, and the other $-1$, if this lattice is to be oriented.

Now homology theory demands that the co-boundary of the co-boundary of a simplex be zero.  This amounts to the matrix equation
\begin{equation}
    \label{bofbiszero}
    \bar{\bound}_{3} \bar{\bound}_{4} = 0,
\end{equation}
with $\bar{\bound}_{3}$ the $N_{2} \times N_{3}$ matrix which has for every 2-simplex the mapping to its co-boundary.  This matrix can be constructed in the following way. Operate on the ordered vertices of each 3-simplex using the boundary operator.  The sign assigned to each 2-simplex from this operation is the matrix element for $\bar{\bound}_{3}$.  The only subtlety here occurs because we work with degenerate triangulations.  In this case the matrix elements of $\bar{\bound}_{p}$ must be specified using the unique identification label of a simplex, since many of the same type of simplex (with different identification labels) will use the same vertices in a degenerate triangulation.  The orientation found for $\bar{\bound}_{4}$ will guarantee Eq.~\eqref{bofbiszero} is zero if the columns of $\bar{\bound}_{3}$ are constructed using the boundary operator.  One then proceeds identically in all the lower sub-simplex matrices.

The full ``co-boundary matrix'' is defined as
\begin{equation}
    \bar{\bound} = 
    \begin{pmatrix}
        0 & 0 & 0 & 0 & 0 \\
        \bar{\bound}_{4} & 0 & 0 & 0 & 0 \\
        0 & \bar{\bound}_{3} & 0 & 0 & 0 \\
        0 & 0 & \bar{\bound}_{2} & 0 & 0 \\
        0 & 0 & 0 & \bar{\bound}_{1} & 0
    \end{pmatrix}
\end{equation}
and the full ``boundary matrix'' is $\bar{\delta} \equiv \bar{\bound}^{T}$.  These are
$N \times N$ matrices with
\begin{equation}
    N = \sum_{i=0}^{4} N_{i},
\end{equation}
the sum of the number of each simplex.  In fact, if one uses the boundary operator in the construction of the $\bar{\bound}_{p}$ matrices, like above, in general
\begin{equation}
    \bar{\bound} \bar{\bound} = 0
\end{equation}
and
\begin{equation}
    \bar{\delta} \bar{\delta} = 0
\end{equation}
follow immediately.  The fact that the matrices constructed on individual lattice geometries satisfy these relations provides a check of the calculation.

\section{Results}
\label{sec:results}

\subsection{Eigenvalues of the K\"{a}hler-Dirac operator}

The \kd operator augmented with a real mass, $K=(\bound - \, \delta) + m_0$, has complex eigenvalues of the form $\pm i \lambda + m_0$ which follows from the real antisymmetric nature of the  massless \kd operator.
The eigenvalues, $\lambda$ can be calculated using the \emph{Hermitian} operator,
\begin{align}
    -\Box + m^{2}_{0} &= (\bound - \, \delta + m_0)(\bound - \, \delta + m_0)^{\dag} \\
    &= \bound \delta + \delta \bound + m^{2}_0.
\end{align}
One simply subtracts off the mass term and square-roots to obtain $|\lambda|$.

Because we are working on a four-sphere there are two exact
zero-modes, one associated with the 4-simplices, and another the 0-simplices \cite{kdferm2018}.
Consider the spectral decomposition of the lattice \kd operator with a non-zero mass:
\begin{align}
	\bar{K} + m_0 &= \sum_{g=1}^{2} m_0 \ket{0, g} \bra{0, g} \\ 
    &+ \sum_{n = 2}^{N} (i\lambda_{n} + m_0) \ket{n} \bra{n},
\end{align}
where $g$ denotes the two-fold degeneracy.  These zero-modes are a result of a remnant $U(1)$ symmetry \cite{Rabin1982,kdferm2018}.  Specifically, the operator
\begin{equation}
	\label{eq:u1sym}
	\Gamma = \bigoplus_{p=0}^{4} (-\mathbb{1})^{p}
\end{equation}
anti-commutes with the massless \kd operator and can be promoted to a $U(1)$ symmetry, $e^{i \theta \Gamma}$, of the massless action.
% This matrix is a similarity transform for the transpose operation on the massless \kd operator,
% \begin{equation}
% 	\Gamma K \Gamma^{-1} = K^{T} = -K.
% \end{equation}
From this it's clear that the two-fold degeneracy comes from transforming eigenstates of $\bar{K}$ by $\Gamma$.

\begin{figure}[t]
    \includegraphics[width=8.6cm]{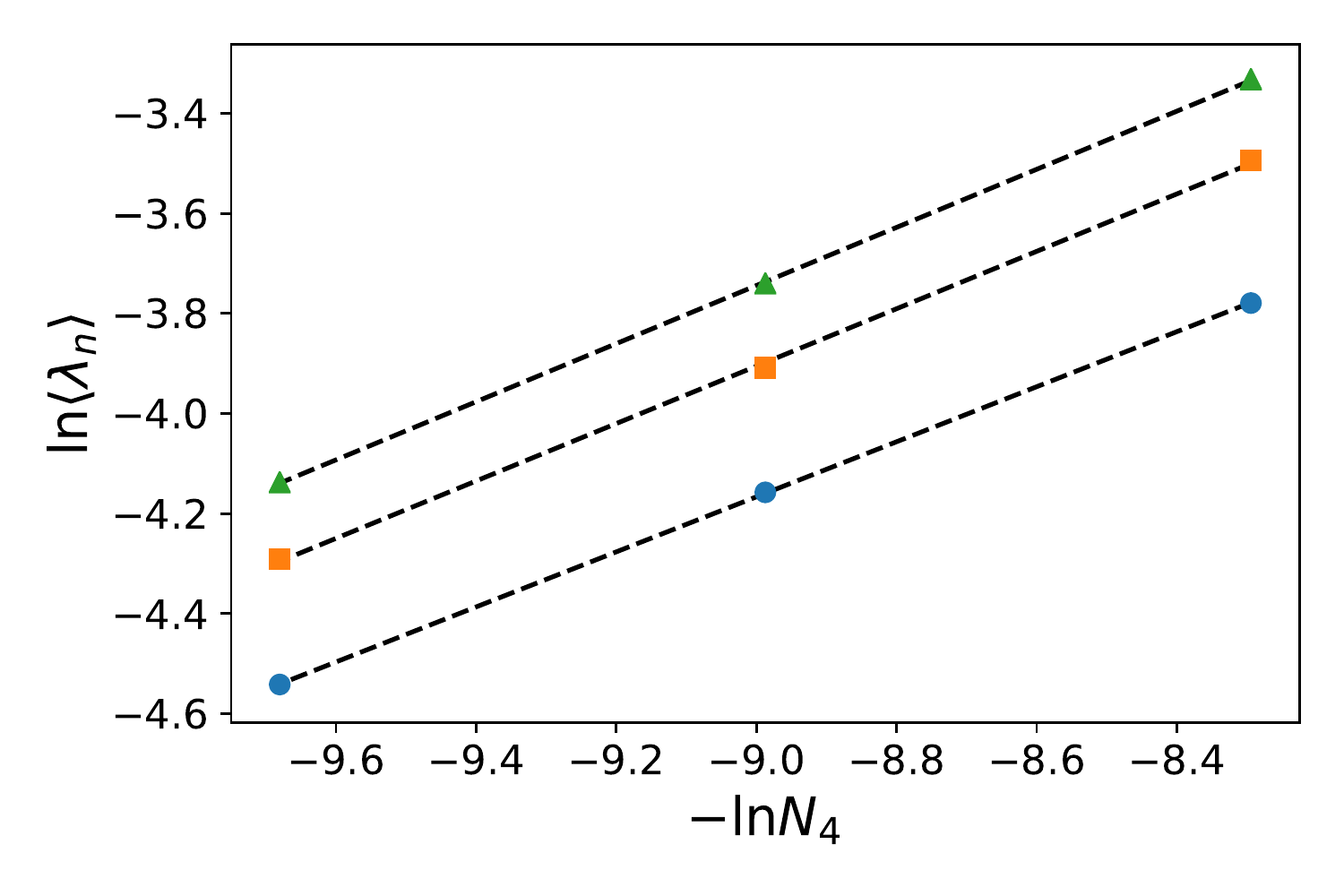}
    \caption{The power law behavior of the first few smallest eigenvalues of the \kd matrix in the system size volume.  Here we show the volume dependence across the 4000, 8000, and 16,000 simplex ensembles at $\beta = 0$.  The dashed lines are linear fits across the three volumes.}
    \label{fig:loglam1}
\end{figure}

\begin{figure}
    \centering
    \includegraphics[width=8.6cm]{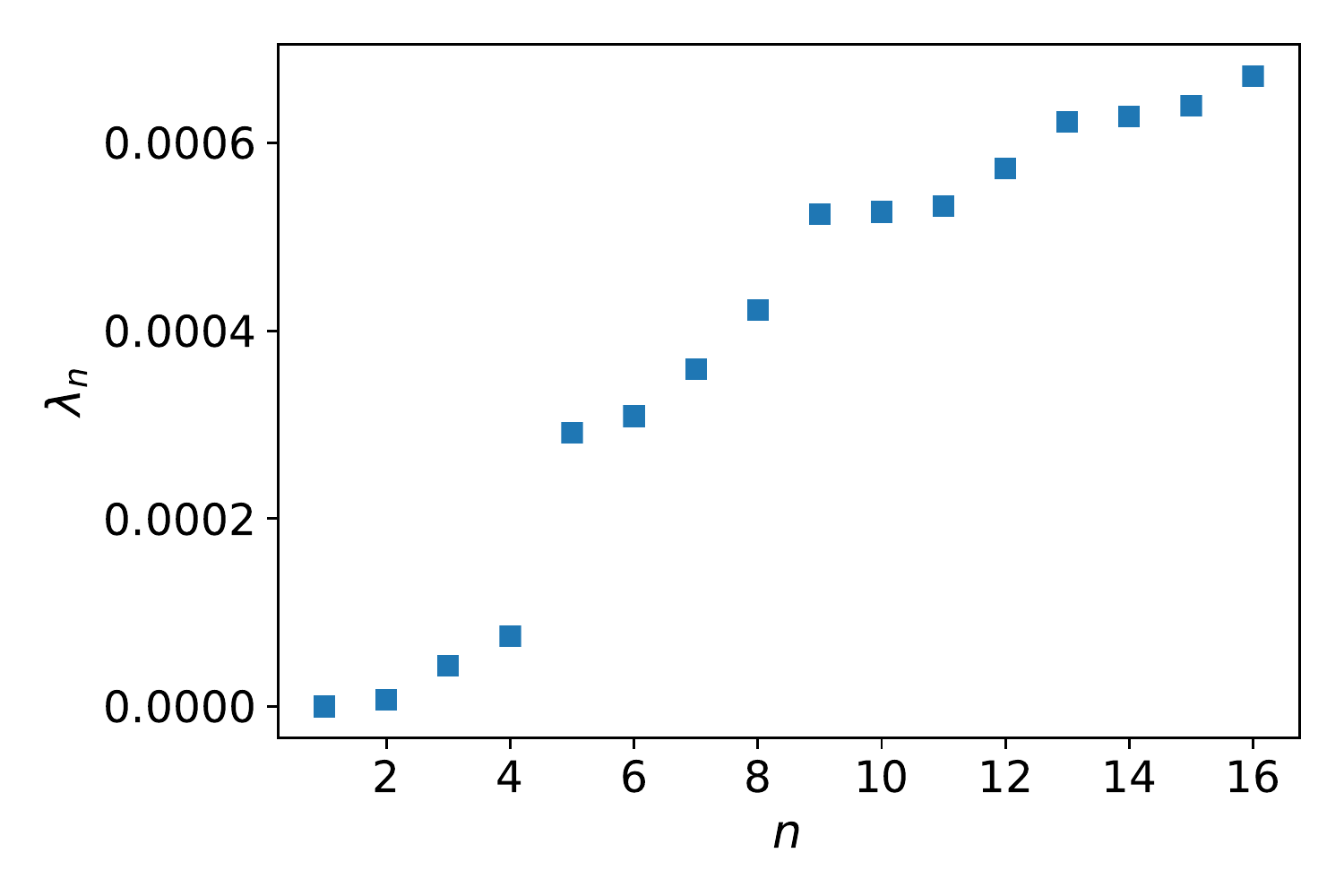}
    \caption{The first 16 eigenvalues of the \kd matrix at infinite-volume extrapolation for the $\beta = 0$ ensembles.}
    \label{fig:degen_eigs}
\end{figure}

We computed the smallest unique 350 eigenvalues of $\bar{K} \bar{K}^{T}$ to extract the eigenvalues of the \kd operator.
Looking at the volume dependence, we can see they follow a power-law behavior.  This can be seen in  Fig.~\ref{fig:loglam1} for the first few eigenvalues.
If we assume a form
\begin{equation}
    \lambda_{n} \simeq \frac{a_n}{N_{4}^{p_{n}}} + b_{n}
\end{equation}
where $a_n$, $p_n$, and $b_n$ are eigenvalue-specific parameters, we can fit and extract the infinite volume value.  To do this we used non-linear least-squares fitting with jack-knife errors.  By preforming this procedure on the first few-hundred eigenvalues we find the eigenvalues re-shuffle themselves in a different order in the infinite volume limit.  In fact, some eigenvalues even extrapolate to negative values, meaning they move from the positive members of $\lambda_n$ to the negative members of $\lambda_n$.  This is acceptable since the same, but negative eigenvalues extrapolate to positive values so the overall number of positive and negative eigenvalues remains the same.  The final re-ordered eigenvalues at infinite volume can be seen in Fig.~\ref{fig:degen_eigs} for the first 16 eigenvalues.
This figure provides some support for the possibility that there exist groups of four-fold degenerate eigenvalues, which the continuum \kd operator has in flat spacetime.  The largest gaps appear after the first four and eight eigenvalues, suggesting that already at coarse coupling the eigenvalues are beginning to arrange themselves in groups of four degenerate values, though finer lattice spacings will be needed to more carefully test this hypothesis.  Plots of this character are reminiscent of lattice QCD calculations of eigenvalues using staggered fermions \cite{Follana2005}, where the fourfold degeneracy is needed to ensure that staggered fermions are in the same universality class as other fermion discretizations.  In our case, this result provides a hint that the infinite-volume, continuum-limit behavior is one in which we recover four copies of Dirac fermions in the small curvature limit, just as in the continuum for \kd fermions.

\subsection{Mesons}

The appearance of approximate four-fold degeneracy of the
eigenvalues of the \kd matrix encourages us to look at other observables which would support the hypothesis that \kd fermions on EDT configurations correspond to four copies of Dirac fermions in the continuum limit.  In this section we consider bound states of \kd fermions. 
By using the propagator for the Klein-Gordon field we can write down explicit formulas for the propagators of \kd fields between simplices.  Let $\bar{K} \equiv (\bar{\bound} - \bar{\delta})$, and $P \equiv (\bar{K}^{2})^{-1}$ with $\bar{K}^2 = \bar{K} \bar{K}^{T}$. Then,
\begin{equation}
    \bar{K}^{-1} = \frac{\bar{K}}{\bar{K}^{2}} = \bar{K} P.
\end{equation}
As a matrix,
\begin{align}
    \nonumber
    (\bar{K} + &m_0)^{-1} \\ &= 
    \begin{pmatrix}
        m_0 P_{4} & -\bar{\delta}_{4} P_{3} & 0 & 0 & 0 \\
        \bar{\bound}_{4} P_{4} & m_0 P_{3} & -\bar{\delta}_{3} P_{2} & 0 & 0 \\
        0 & \bar{\bound}_{3} P_{3} & m_0 P_{2} & -\bar{\delta}_{2} P_{1} & 0 \\
        0 & 0 & \bar{\bound}_{2} P_{2} & m_0 P_{1} & -\bar{\delta}_{1} P_{0} \\
        0 & 0 & 0 & \bar{\bound}_{1} P_{1} & m_0 P_{0}
    \end{pmatrix}
\end{align}
where $P_{n}$ is the $n$-simplex block of $P$.  From this construction we can see that the \kd propagators between simplices of the same type are simply the Klein-Gordon propagators scaled by $m_0$.  There are non-trivial propagators between simplices of different types on the off-diagonals.  We see that a propagator from a $p$-simplex to a $p \pm 1$-simplex is built from linear combinations of propagators from a $p$-simplex to $p$-simplices in the (co)boundary of the $p \pm 1$-simplex.  For example, the propagator between a 4-simplex and a 3-simplex is the difference (because they are oriented) of the propagators from the 4-simplex to the two co-boundary 4-simplices of the 3-simplex.  With these matrix elements, and a definition of distance between simplices, we can compute propagators.

Let us now turn to a definition of distance for the lattice propagators.  Apart from the 0- and 4-simplex propagators, a clear definition of distance between simplices is not available.  Some ideas are:
\begin{enumerate}
	\item Use the non-zero, off-diagonal indices of the matrix $\bar{\bound}_{p} \bar{\delta}_{p}$ to identify neighbors.  These are all neighboring simplices which share a $p+1$-simplex.  This seems natural on the dual lattice.
    \item Use the non-zero, off-diagonal indices of the matrix $\bar{\delta}_{p} \bar{\bound}_{p}$ to identify neighbors.  These are all neighboring simplices that share a $p-1$-simplex.  This seems natural on the original lattice.
    \item Use the non-zero, off-diagonal indices of the matrix $\bar{\bound}_{p} \bar{\delta}_{p} + \bar{\delta}_{p} \bar{\bound}_{p}$ to identify neighbors.  These are all neighboring simplices which share a $p+1$-simplex, or a $p-1$-simplex, but not both.  This option is justified by the fact that it lets the lattice Laplacian define nearest neighbors.
    \item Smear the source and sink at each 4-simplex.  Assign a distance based on the dual lattice edges (tetrahedra) and average the source-to-sink propagators between 4-simplices.  This is simple and easy to implement.
\end{enumerate}
Items 1-3 require one to ``fan-out'' from the single source sub-simplex in a consistent fashion, which is possible but difficult.  The results of this work use option 4, which is simple to implement and shows a cleaner signal than others that were looked at.  Ultimately only the short distance physics should be affected by smearing.

Given that there are two zero-modes in the eigenvalue spectrum it is natural to look at quartic operators of fermion fields, since the two factors of mass will cancel the two factors from the fermion determinant in the dynamical theory in the chiral limit \cite{Creutz2008,kdferm2018}.
One simple case to consider is the analog of the pion for each of the possible simplex-sectors,
\begin{align}
	\langle \bar{\omega}_{x} \Gamma \, \omega_{x}
    \bar{\omega}_{y} \Gamma \, \omega_{y} \rangle &\rightarrow \langle \omega_{y} \bar{\omega}_{x} \rangle \langle (\omega_{y} \bar{\omega}_{x})^{\dag} \rangle \\
    &= (\bar{K} + m_0)^{-1}_{yx} ((\bar{K} + m_0)^{-1}_{yx})^{*}, \\
    &\equiv \langle \pi(x) \pi(y) \rangle
\end{align}
which is built from two fermion propagators \footnote{We are ignoring possible disconnected contributions which amounts
to working with 2 flavors of \kd field}.
It is straightforward to construct pion-like meson operators in the nine main sectors of the inverse \kd matrix.  These nine sectors are the $p$-simplex to $p$-simplex type mesons located on the diagonal, and the $p$-simplex to $p \pm 1$-simplex type mesons on the off-diagonal.

\begin{figure}
    \centering
    \includegraphics[width=8.6cm]{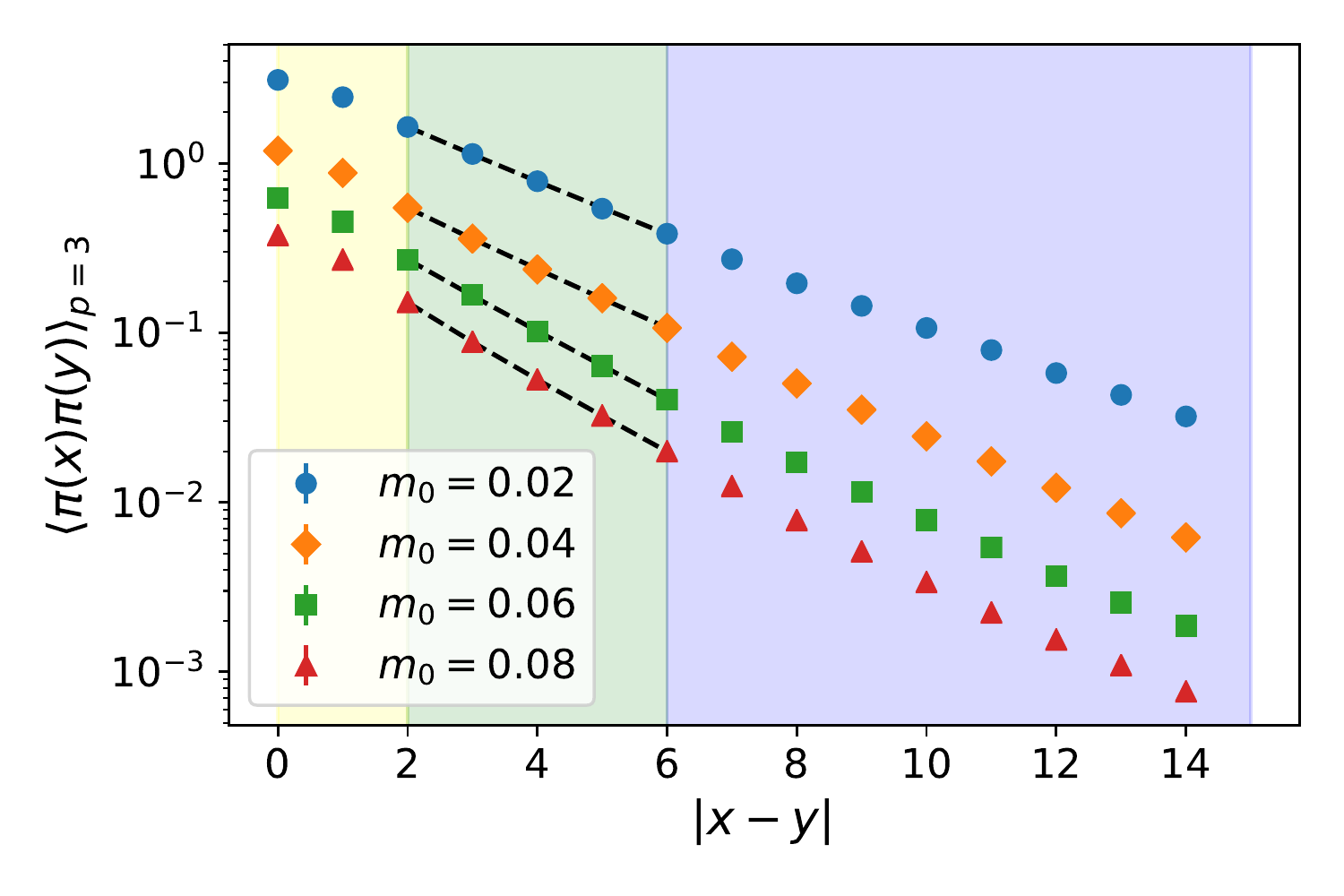}
    \caption{Meson propagators for the 3-simplex to 3-simplex block for four different masses, along with their fits in black using Eq.~\eqref{eq:powerexp}.  Here the three different distance regions are shown (left to right) as yellow, green, and blue.  This data is from the $\beta = -0.8$, 8000 simplex ensemble.}
    \label{fig:tet-meson}
\end{figure}

We primarily consider the $p$-simplex-to-$p$-simplex mesons.  From the meson propagators one can extract a mass.  The meson propagators, as a function of distance, can typically be divided into three regions: short distances, intermediate distances, and long distances. At short distances the signal can be dominated by discretization effects from the finite lattice spacing, while
at long distances there are discretization effects that might be attributed to baby universes that branch off the mother universe.  These baby universes can be quite long at coarse lattice spacings, though their cross-section is of order the lattice cut-off.  Although this baby universe contamination appears to vanish in the continuum limit, it is important to keep in mind that it can lead to significant deviations from semiclassical behavior at long distances on coarse lattices.  We use the lattice distance where the semi-classical approximation for a de~Sitter spacetime broke down to determine a long distance cutoff in our fits \cite{Laiho:2016nlp}.
Similarly at very short distances the functional form of the propagator is contaminated by discretization
effects and excited states, therefore, one must look for a window of intermediate distance scales that are in the semiclassical scaling regime and extract a meson mass from fits to that range.  This range is typically between a few lattice spacings, and $\sim 15$ lattice spacings depending on the ensemble.  As an example, the 3-simplex meson propagator for the $\beta = -0.8$, 8000 simplex ensemble is shown in Fig.~\ref{fig:tet-meson} for four different masses along with the approximate boundaries for the three distance regions.
We use two different fitting forms depending on the lattice spacing.  For coarser lattices we use an exponential fit,
\begin{equation}
    \langle \pi(x) \pi(y) \rangle \simeq a \exp{(-m |x-y|)},
\end{equation}
while for finer lattices we use
\begin{equation}
\label{eq:powerexp}
    \langle \pi(x) \pi(y) \rangle \simeq
    a \exp{(-m |x-y|)}/|x-y|^b,
\end{equation}
with $a$, $b$, and $m$ fit parameters.  The latter fit form includes power law dependence, which is expected when the mesons propagate on a curved background \cite{deBakker:1996qf}.  The rationale for the two fit forms is that the finer lattices also have smaller physical volumes, and thus the effects of curvature are expected to be more evident, which appears to be the case.

\begin{figure}
    \centering
    \includegraphics[width=8.6cm]{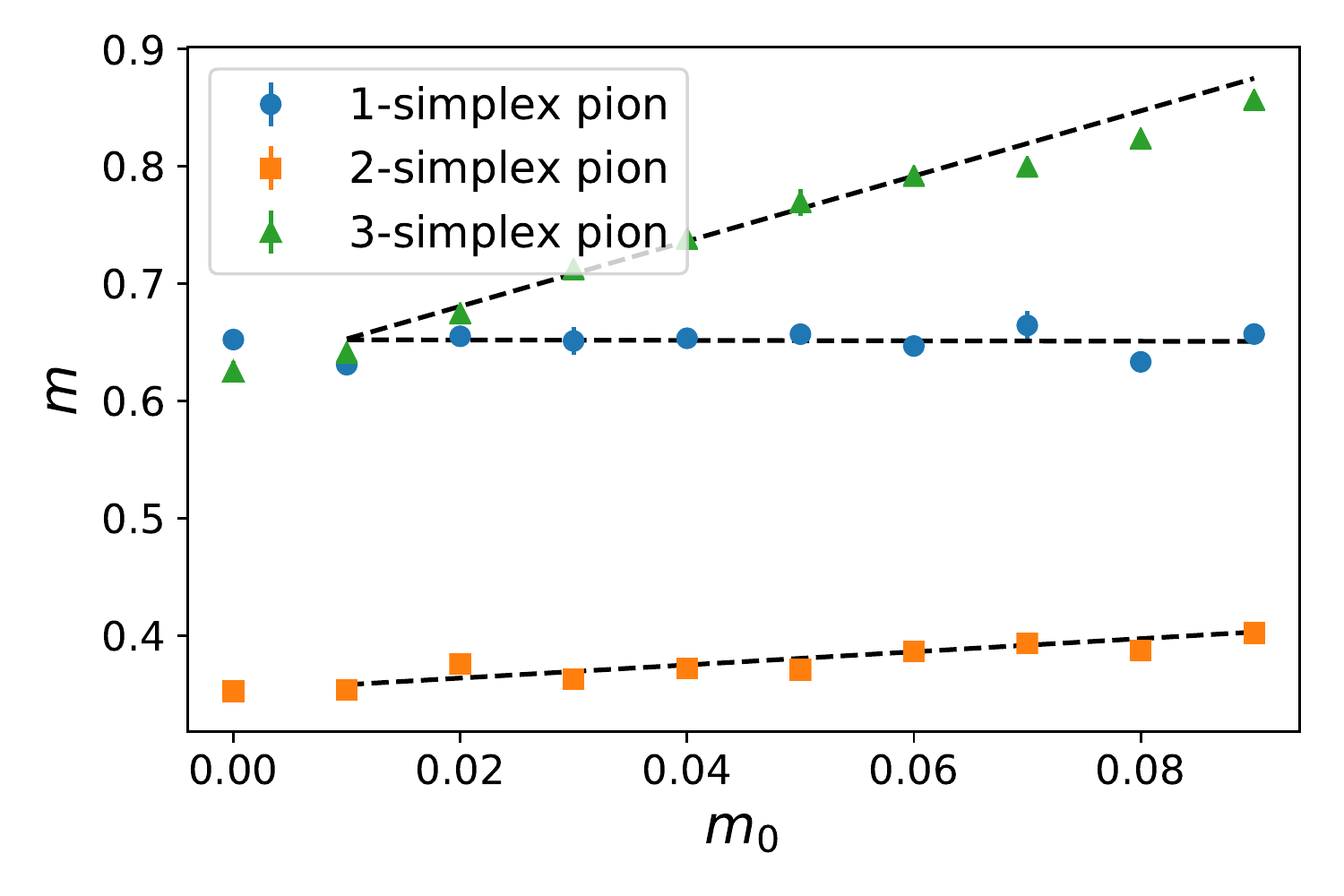}
    \caption{The chiral extrapolation for the 1-, 2-, and 3-simplex mesons.  Here the dashed lines are linear fits, and the furthest left data points are the $m_0 \rightarrow 0$ extrapolation.  These masses are from the 4000 simplex, $\beta = 0$ ensemble.}
    \label{fig:chi123extra}
\end{figure}

We find that, except for the 0-simplex meson, the other mesons tend to have much larger masses than the input bare quark mass.  We expect the 0-simplex meson to vanish in the chiral limit since there is a zero-mode associated with that sector, and we also expect that the 4-simplex meson mass should tend to zero in the chiral limit since there is also a zero mode associated with that sector.  The 0-simplex meson shows every sign of vanishing in the chiral limit.  The 4-simplex meson is more massive for larger input masses, but does indeed tend to zero quickly in the small mass regime.  The intermediate simplex mesons clearly remain massive in the chiral limit as can be seen in Fig.~\ref{fig:chi123extra}, however as we discuss below, all the mesons exhibit behavior consistent with $m \rightarrow 0$ in the chiral limit if we also take the continuum limit.

\begin{figure}
    \centering
    \includegraphics[width=8.6cm]{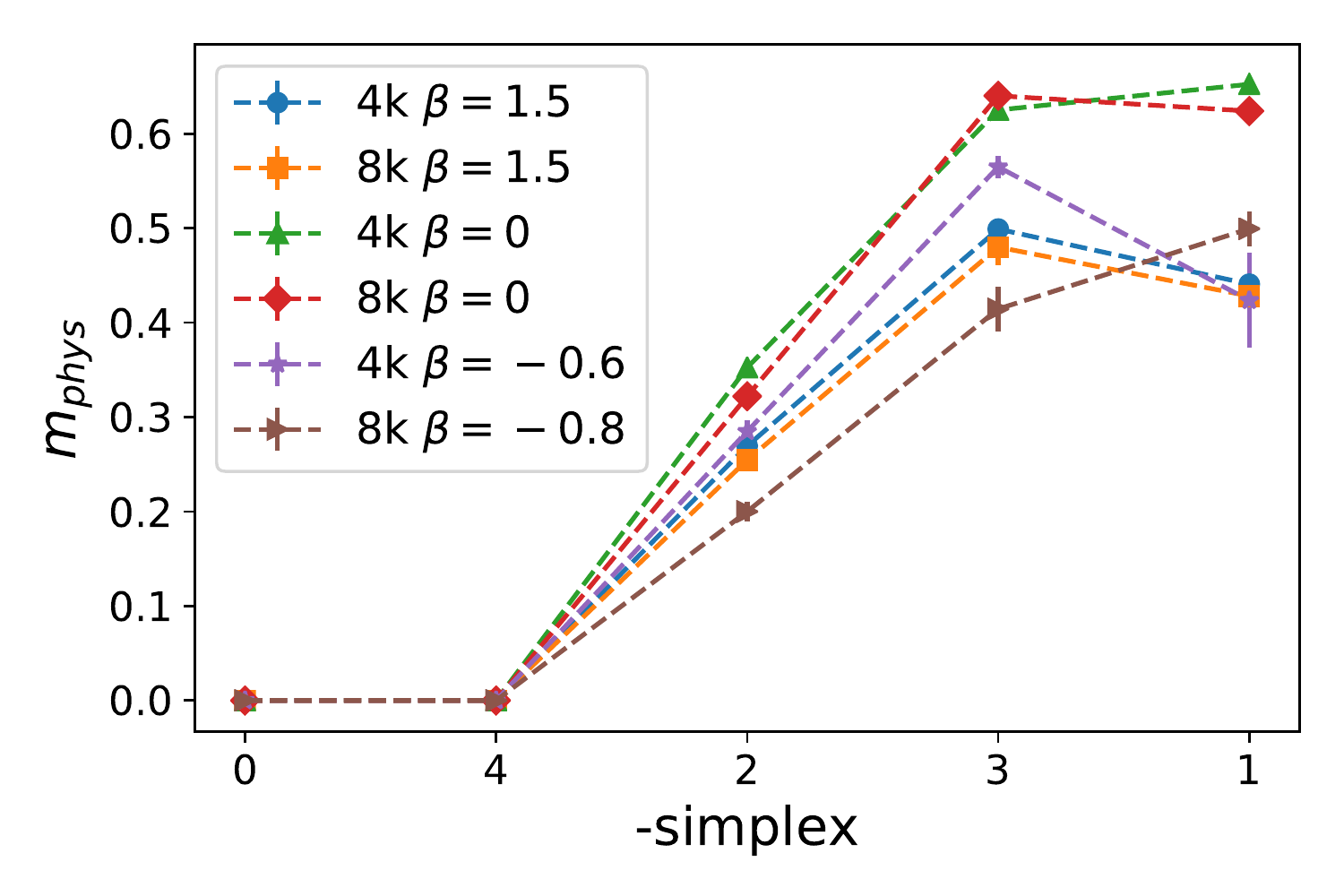}
    \caption{The physical masses of the meson propagators between like simplices in the chiral limit for a variety of ensembles.  The 0- and 4-simplex masses tend to zero in the chiral limit because of the inherent zero modes.  The other simplex mesons remain massive in the chiral limit; however, as the lattice become finer their masses tend to get smaller.  The coarsest lattices do not follow this trend, but the expected scaling should break down for sufficiently large lattice spacing.  The relative lattice spacing is set by the fiducial lattice spacing at $\beta = 0$.}
    \label{fig:pion_clim}
\end{figure}
\begin{figure}
    \centering
    \includegraphics[width=8.6cm]{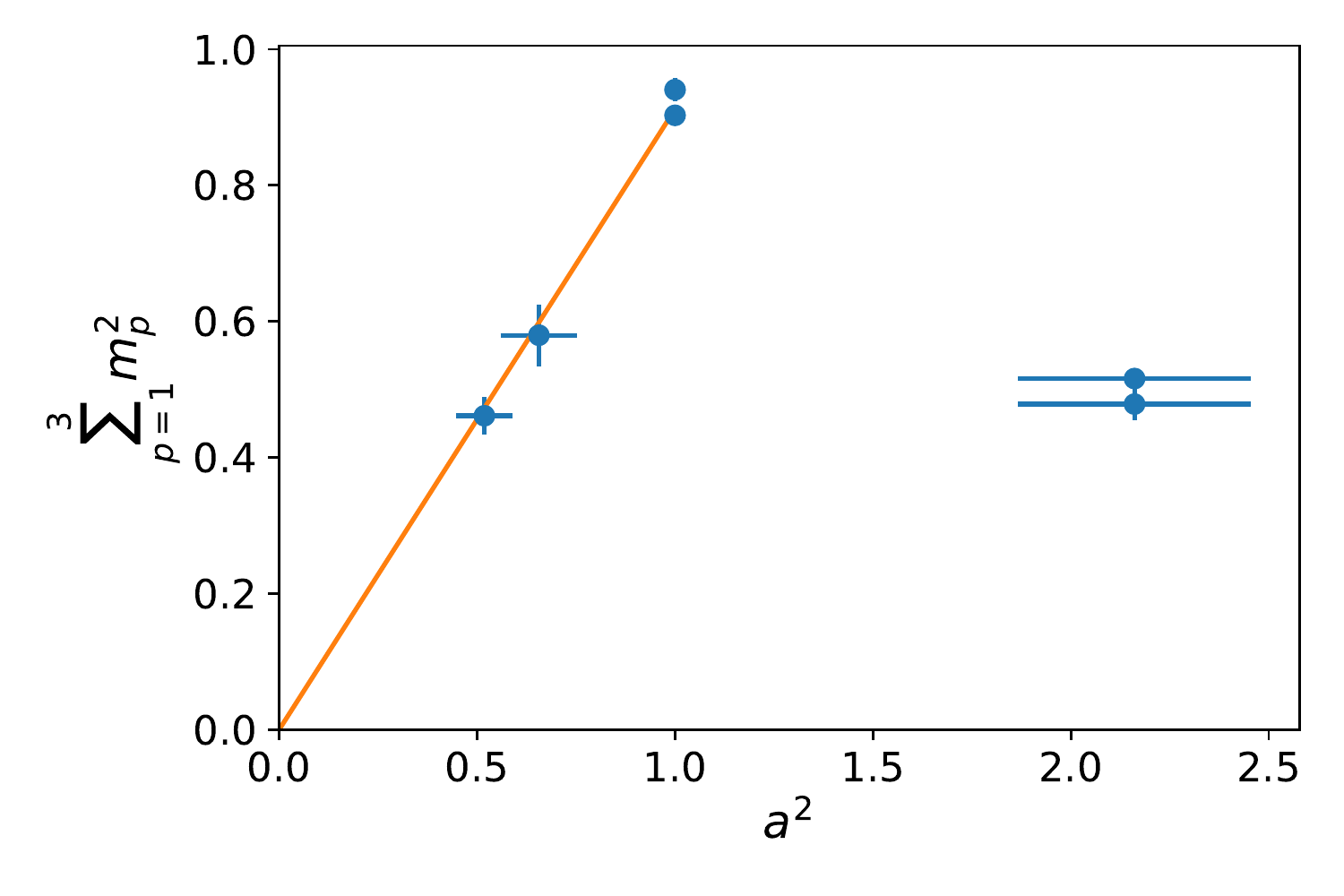}
    \caption{The sum of the squares of the meson masses for the \\ 1-, 2-, and 3-simplex mesons versus the squared relative lattice spacing.  At coarse lattice spacing there doesn't seem to be a particular trend; however, at finer lattices these masses trend to zero.  The line is fixed at zero and the lower $a_{\rm rel}^2=1$ point and is meant to guide the eye.}
    \label{fig:sqsum}
\end{figure}

Figure~\ref{fig:pion_clim} shows the masses of each of the mesons in the chiral limit for a few ensembles representing different lattice spacings.  Here the masses are expressed in physical units using a fiducial lattice spacing set to that at $\beta = 0$.  In this figure we see that the masses associated with the finest lattice spacings are trending smaller.  This would be in line with completely degenerate fermion masses in the continuum, infinite-volume limit.  Indeed if we look at the sum of the squares of the masses for the 1-, 2-, and \\ 3-simplex mesons as a function of squared lattice spacing we find these masses are tending to zero linearly.  This is shown in Fig.~\ref{fig:sqsum}.
Here a line to guide the eye has been drawn, fixed at zero and ending at the lower $a_{\rm rel}^2=1$ point.
Note that the coarsest lattices do not follow this linear trend.  This might be expected, since for sufficiently large lattice spacings the scaling behavior in the fixed point regime should break down.

\subsection{Condensates}

Another important aspect that \kd fermions should satisfy if they are to model realistic continuum fermions is that chiral symmetry should not be broken at order the Plank scale.  Here we investigate this possibility when \kd fermions are coupled to strongly-coupled gravity.
We consider the diagonal of the inverse \kd matrix, as the trace of this matrix corresponds
to the vacuum expectation value (VEV) of the fermion bilinear. 
Here we define the bilinear VEV as
\begin{align}
    \label{eq:kd_bilinear}
	\langle \bar{\omega} \omega \rangle &= \frac{1}{N_{4}} \sum_{x} (\bar{K} + m_0)^{-1}_{xx}
    \\
    &= \frac{1}{N_{4}} \left( \frac{2}{m_0} +  2m_0 \sum_{n=2}^{N/2} \frac{1}{\lambda_{n}^{2} + m^{2}_0} \right) 
\end{align}
on the lattice using the \kd matrix.  Using stochastic $Z_{2}$ noise \cite{Dong1994}, it is relatively simple to extract the diagonal elements of the inverse \kd matrix.  Using a small number of configurations within each ensemble along with small ensembles of $Z_{2}$ noise yields relatively clean results. The presence of a condensate would be signaled
by this observable tending to a finite limit as $m_0\to 0$ and $N_{4}\to\infty$.
\begin{figure}[t]
	\includegraphics[width=8.6cm]{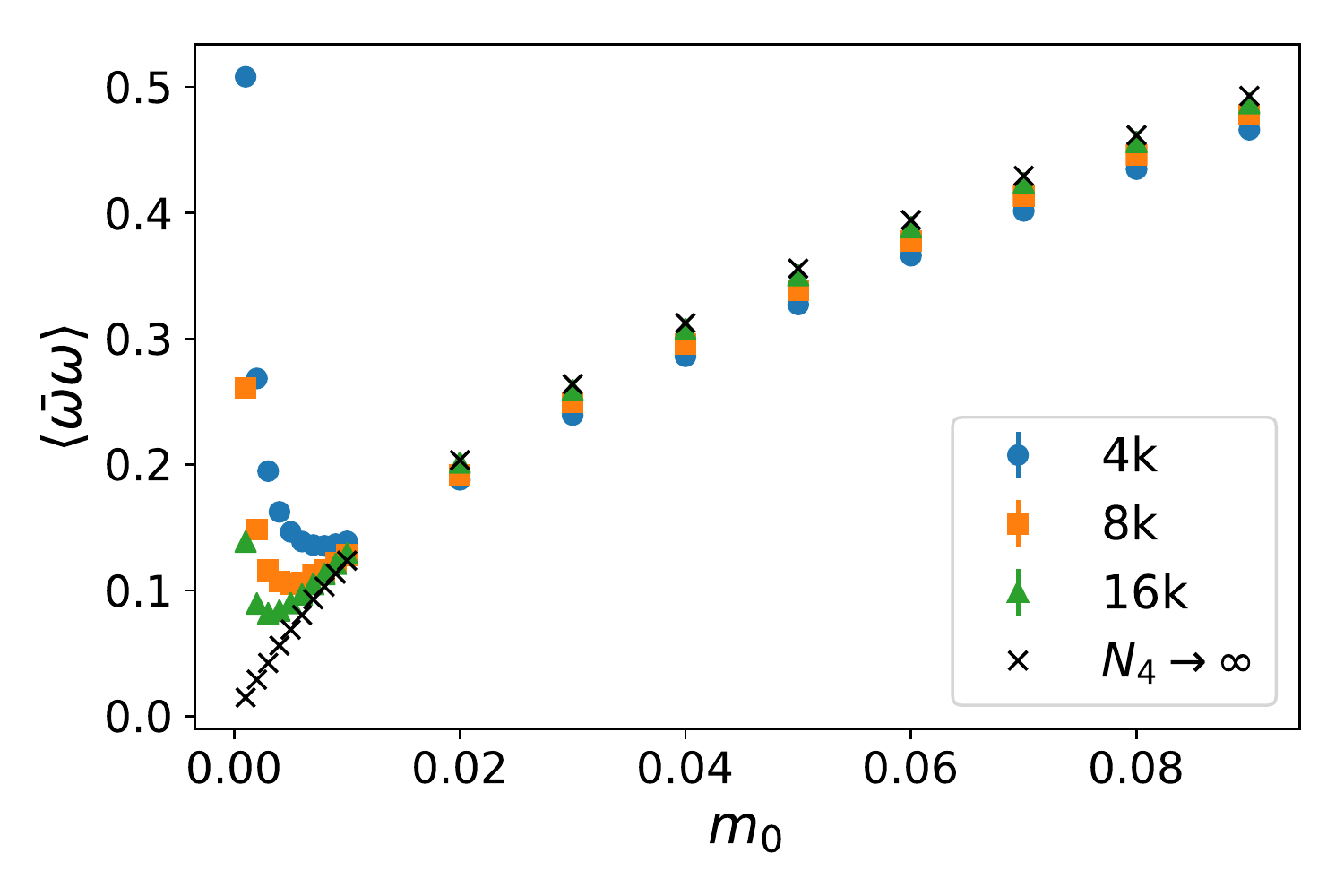}
    \caption{The bilinear condensate for the \kd matrix at three different volumes and the extrapolation to infinite volume.  The finite volume curves show that the condensate diverges for sufficiently small values of $m_0$.  Taking the infinite volume limit and then taking the chiral limit we see that the condensate vanishes as $m_0 \rightarrow 0$.  The extrapolation to zero in the chiral limit indicates the $U(1)$ symmetry associated with Eq.~\eqref{eq:u1sym} is not spontaneously broken.  These calculations were done at $\beta = 0$.}
    \label{fig:pbp-wzm}
\end{figure}
Figure~\ref{fig:pbp-wzm} shows the bilinear condensate across three volumes at a fixed value of $\beta = 0$, as well as the infinite volume extrapolation.  At finite volume one can see the $1/m_0$ divergence at sufficiently small mass predicted by the form of Eq.~\eqref{eq:kd_bilinear}.  Following Refs.~\cite{Banks1980,Leutwyler1992}, the removal of the zero-modes can be justified and one can extract the chiral limit from the finite volume condensate; however, ideally one would first take the $N_{4} \rightarrow \infty$ limit and then the $m_0 \rightarrow 0$ limit.  We tried both removing the zero-modes at finite volume, as well as taking the $N_{4} \rightarrow \infty$, $m_0 \rightarrow 0$ limit.  The two procedures agree.  Here we show results from the later.  We use a fitting form,
\begin{equation}
    \langle \bar{\omega} \omega \rangle \simeq
    \frac{a}{N_4} + b,
\end{equation}
where $a$ and $b$ are free parameters, to extract the infinite volume condensate.
The extrapolation to zero for the bilinear condensate in the infinite volume, chiral limit indicates that the remnant $U(1)$ symmetry associated with $\Gamma$ is not spontaneously broken in the limit of vanishing
bare mass (what is usually termed the chiral limit) \cite{Leutwyler1992,Espriu1986}.

Yet another possible condensate is a four-fermion condensate defined on the lattice as
\begin{equation}
	\langle \bar{\omega} \omega \bar{\omega} \omega \rangle = \frac{1}{N_4} \sum_{x} ((\bar{K} + m_0)^{-1}_{xx})^{2}.
\end{equation}
Similar to the bilinear, this quantity diverges for small $m_0$, however the divergence goes like $1/m^{2}_0$ in this case, as opposed to $1/m_0$.  This term should arise naturally automatically in simulations
with dynamical \kd fermions.  It is associated with an anomalous breaking of the $U(1)$ symmetry on
manifolds with non-zero Euler number. However, the condensates arising from the anomaly
vanish like the inverse volume and are far too small to be seen in our simulations.   This is discussed in detail in Ref.~\cite{kdferm2018}.

\section{Conclusion}
\label{sec:conc}

We have investigated the coupling of \kd fermions to dynamical triangulations in the quenched approximation.  Here only gravity is dynamical, i.e. we neglect the back-reaction from the matter sector.  The lattice \kd matrix is constructed using results from homology theory, and this construction allows us to avoid introducing a spin connection and a vielbein explicitly within the formulation.  Instead, one can simply rely on the lattice analogues of the exterior derivative and its adjoint, the so-called coboundary and boundary operators. The use of \kd fermions has a further advantage - they exhibit an
exact $U(1)$ symmetry at vanishing bare mass on any triangulation and hence fermion masses only
receive multiplicative renormalization.

Using this construction we study the eigenvalues, meson propagators and bilinear fermion condensate of the \kd matrix.  The observed eigenvalues matched expectations from the topology of the lattice configurations, namely the number of zero-modes and their respective simplex sectors.  In the thermodynamic limit there is a hint that the eigenvalues are clumping into degenerate groups of four.  This is consistent with the scenario in which \kd fermions become four copies of Dirac fermions in the small curvature limit, as expected in the continuum.  However, we are only able to take the infinite volume limit for a single lattice spacing and additional volumes at finer lattice spacings are necessary to make this study more conclusive.

We also considered meson propagators between like simplices.  Here we looked across a range of relative lattice spacings and compared meson masses in physical units in the limit of small bare masses (the chiral limit).  We found the 0- and 4-simplex mesons tended towards zero mass in the chiral limit while the intermediate simplex mesons had finite mass.  However, as one moves along the critical line to finer lattices we see that the sum of the squared splittings extrapolates to zero and that the extrapolation is linear in the lattice spacing squared.  This is further evidence that the four species of \kd fermions become degenerate in the continuum (infinite volume) limit.  Taken together, these results provide non-trivial evidence that we recover the correct long-distance behavior of \kd fermions in the continuum.  This provides evidence that fermions see the emergence of four-dimensional, smooth Euclidean space with small curvature at the same
point where fits of the geometry indicate a semi-classical de Sitter space.
The fact that we see discretization effects that are strongly reminiscent of staggered fermions in QCD and that these effects appear to vanish in the continuum lends further support to the fact that the continuum limit actually exists, and thus bolsters the case for the asymptotic safety of quantum gravity as realized by dynamical triangulations.

We also looked at the bilinear condensate associated with the inverse of the \kd operator and found that it matches our expectations based on our study of the eigenvalues.  We find a $1/m_0$ divergence for sufficiently small $m_0$ at finite volume, but upon taking the infinite volume limit the condensate tends to zero.  This indicates a lack of breaking of the remnant $U(1)$  symmetry.  Although further study at finer lattice spacings is needed to solidify these results, our work suggests that there is no spontaneous chiral symmetry breaking of order the Planck scale, so that fermion bound states do not acquire an unacceptably large mass due to strongly coupled gravity in our model.  Similar results were found in functional renormalization group investigations of fermions coupled to quantum gravity \cite{Eichhorn:2011pc}.

Finally, it would be interesting to un-quench our simulations of \kd fermions coupled to gravity.  Then we could compare our current results with identical quantities calculated including the back-reaction of the matter fields, thus eliminating a short-coming of the present work.  We are currently investigating various algorithms towards this end.

\begin{acknowledgments}
The authors thank Scott Bassler for his assistance in providing the 8000 4-simplex, $\beta=1.5$ configurations used in this work, and Fleur Versteegen for valuable conversations.
This work is supported in part by the U.S.~Department of Energy, Office of Science, Office of High Energy Physics, under Award Number DE-SC0009998.
Computations for this work were carried out in part on facilities of the USQCD Collaboration, which are funded by the Office of Science of the U.S. Department of Energy, on the Darwin Supercomputer as part of STFC's DiRAC facility jointly funded by STFC, BIS, and the Universities of Cambridge and Glasgow, and on the Syracuse University condor cluster, which is funded by the National Science Foundation under Grant No. ACI-1341006.
\end{acknowledgments}

% \bibliography{kdbib}
%

\end{document}